\documentclass[preprint,3p,number,a4paper,sort&compress]{elsarticle}
\usepackage{graphicx}
\usepackage{float}
\usepackage{mathptmx, courier, pifont}
\usepackage[scaled=0.92]{helvet}
\usepackage[T1]{fontenc}
\usepackage{textcomp}
\usepackage{color}
\usepackage[dvipsnames]{xcolor}
\usepackage[colorlinks=true,urlcolor=Blue,linkcolor=Blue]{hyperref}
\usepackage[all]{hypcap}
\usepackage{lineno,hyperref}
\usepackage{multirow}
\usepackage{longtable}
\usepackage{bigstrut}
\usepackage{booktabs}
\usepackage{threeparttable}
\usepackage{array}
\usepackage{amssymb}
\usepackage{amsmath}
\usepackage{ulem}
\modulolinenumbers[1]
\begin{document}
%\preprint{APS/123-QED}
\begin{frontmatter}
\title{Determination of luminosity for in-ring reactions: A new approach for the low-energy domain}
\cortext[mycorrespondingauthor]{Corresponding authors}
\author[imp,gsi]{Y.M. Xing}
\author[gsi]{J. Glorius\corref{mycorrespondingauthor}}
\ead{J.Glorius@gsi.de}
\author[gsi]{L. Varga}
\author[guf]{L. Bott}
\author[gsi,jlu]{C. Brandau}
\author[guf]{B. Br\"uckner}
\author[imp,gsi]{R.J. Chen}
\author[imp]{X. Chen}
\author[bau]{S. Dababneh}
\author[edin]{T. Davinson}
\author[guf]{P. Erbacher}
\author[guf]{S. Fiebiger}
\author[gsi]{T. Ga{\ss}ner}
\author[guf]{K. G\"obel}
\author[guf]{M. Groothuis}
\author[gsi]{A. Gumberidze}
\author[atomki]{G. Gy\"urky}
\author[gsi]{M. Heil}
\author[gsi]{R. Hess}
\author[guf]{R. Hensch}
\author[guf]{P. Hillmann}
\author[gsi]{P.-M. Hillenbrand}
\author[guf]{O. Hinrichs}
\author[cenbg]{B. Jurado}
\author[guf]{T. Kausch}
\author[gsi,guf]{A. Khodaparast}
\author[guf]{T. Kisselbach}
\author[guf]{N. Klapper}
\author[gsi]{C. Kozhuharov}
\author[guf]{D. Kurtulgil}
\author[anu]{G. Lane}
\author[guf]{C. Langer}
\author[edin]{C. Lederer-Woods}
\author[gsi]{M. Lestinsky}
\author[gsi]{S. Litvinov}
\author[gsi]{Yu.A. Litvinov\corref{mycorrespondingauthor}}
\ead{y.litvinov@gsi.de}
\author[tud,gsi]{B. L\"oher}
\author[gsi]{N. Petridis}
\author[gsi]{U. Popp}
\author[anu]{M. Reed}
\author[guf]{R. Reifarth}
\author[gsi]{M. S. Sanjari}
\author[gsi]{H. Simon}
\author[guf]{Z. Slavkovsk\'a}
\author[gsi]{U. Spillmann}
\author[gsi]{M. Steck}
\author[gsi,hij]{T. St\"ohlker}
\author[guf]{J. Stumm}
\author[atomki]{T. Sz\"ucs}
\author[guf]{T. T. Nguyen}
\author[guf]{A. Taremi Zadeh}
\author[guf]{B. Thomas}
\author[spbu]{S. Yu. Torilov}
\author[gsi,tud]{H. T\"ornqvist}
\author[gsi,jlu]{C. Trageser}
\author[gsi]{S. Trotsenko}
\author[guf]{M. Volknandt}
\author[imp]{M. Wang}
\author[guf]{M. Weigand}
\author[guf]{C. Wolf}
\author[edin]{P. J. Woods}
\author[imp]{Y.H. Zhang}
\author[imp]{X.H. Zhou}
\address[imp] {Key Laboratory of High Precision Nuclear Spectroscopy and Center for Nuclear Matter Science, Institute of Modern Physics, Chinese Academy of Sciences, Lanzhou 730000, China}
\address[gsi] {GSI Helmholtzzentrum f{\"u}r Schwerionenforschung, Planckstra{\ss }e 1, Darmstadt 64291, Germany}
\address[guf] {Goethe Universit\"at, Theodor-W.-Adorno-Square 1, Frankfurt am Main 60323, Germany}
\address[jlu] {Justus-Liebig Universit\"at, Ludwigstra{\ss }e 23, Gie{\ss}en 35390, Germany}
%\address[mpi] {Max-Planck-Institut f\"ur Kernphysik (MPIK), Saupfercheckweg 1, Heidelberg 69117, Germany}
\address[bau] {Al-Balqa Applied University, P.O. Box, Salt 19117, Jordan}
\address[edin] {University of Edinburgh, South Bridge, Edinburgh EH8 9YL, United Kingdom}
\address[atomki] {Institute for Nuclear Research (MTA Atomki), Bem t\'er 18/c, Debrecen 4026, Hungary}
\address[cenbg] {CENBG, CNRS-IN2P3, Rue du Solarium 19, Gradignan 33170, France}
\address[anu] {Australian National University, Canberra ACT 2600, Australia}
\address[tud] {Technische Universit\"at Darmstadt, Karolinenpl. 5, Darmstadt 64289, Germany}
\address[hij] {Helmholtz-Insitut Jena, Fr\"obelstieg 3, Jena 07743, Germany}
\address[spbu] {St. Petersburg State University, Lieutenant Schmidt emb., 11/2, St. Petersburg 199034, Russia}
%\address[ptb] {Physikalisch-Technische Bundesanstalt, Bundesallee 100, Braunschweig 38116, Germany}
%\address[tub] {Technische Universit\"at Braunschweig, Universit\"atspl. 2, Braunschweig 38106, Germany}
%\edin
\date{\today}

\begin{abstract}
%Background, Purpose, Methods, Results, and Conclusions
Luminosity is a measure of the colliding frequency between beam and target and it is a crucial parameter for the measurement of absolute values, such as reaction cross sections.
In this paper, we make use of experimental data from the ESR storage ring to demonstrate that the luminosity can be precisely determined by modelling the measured Rutherford scattering distribution.
The obtained results are in good agreement with an independent measurement based on the x-ray normalization method.
Our new method provides an alternative way to precisely measure the luminosity in low-energy stored-beam configurations. This can be of great value in particular in dedicated low-energy storage rings where established methods are difficult or impossible to apply.
\end{abstract}

\begin{keyword}
\texttt ~luminosity\sep Rutherford scattering\sep storage ring\sep beam\sep gas target\sep reaction

\end{keyword}
%\MSC[2010] 00-01\sep  99-00
%\pacs {21.10.Dr, 27.40.+z, 29.20.db}
%\maketitle
\end{frontmatter}
\section{Introduction}
%[words] beam-introduced minimum chi-square method  the X-ray transition chi¨Csquare minimization  the geometry of beam-target overlap  located on the high energy tail of the Gamow window
Luminosity is a key parameter used in the experiments for absolute cross section measurement~\cite{Grafstrom2015}.
%With well-known luminosity, one can study cross sections which are of great interest for astrophysics~\cite{Brune2015} or other studies~\cite{Gumberidze2015}.
To directly measure luminosity, detailed knowledge of beam intensity and target density is required.
Beam intensity can be precisely measured by a calibrated current transformer in a storage ring~\cite{Stohlker1995} or a beam calorimeter for stopped beams~\cite{Cavanna2014}, while the measurement of the target density depends on experimental conditions.
For a solid target, the thickness can be well estimated to a precision below 5\%~\cite{Stohlker1995,Scott2012}.
However, to precisely determine the effective gas target density, one has to precisely measure the temperature and pressure as a function of the position inside the gas target~\cite{Cavanna2014}.
%In order to determine the effective target density, the temperature and pressure of the target gas have to be known as a function of the position inside the gas target.
Sometimes, heat transfer from the intense ion beam which may influence the density of the target gas along the beam path should also be considered~\cite{Marta2006,Bemmerer2006,Cavanna2014}.
Particularly, for the case of using an internal gas target in a storage ring, detailed knowledge of the target and beam profiles as well as the beam-target overlap~\cite{Shao2016,Eichler2007} is additionally required and an overall uncertainty of 30\% may be assumed~\cite{Eichler2007}.

To remove this large luminosity uncertainty in reactions with an internal gas target in a storage ring, indirect methods are widely applied.
For example, the beam energy loss in the gas target has been used for the thickness determination of the target as well as for the luminosity~\cite{Shao2016,Stein2008}, where a 5\% precision was generally obtained.
However, this precision highly depends on the accuracy of the measured revolution frequency shifts due to the beam energy loss.
If the beam lifetime is too short for a reliable frequency shift measurement, the application of this method is limited.
Alternatively, the luminosity can be obtained from the analysis of a reference reaction with a well-known cross section.
It is based on the precondition that the reference cross section is better known than the cross section to be measured.
However, there are often difficulties in finding a suitable reference reaction.
%This allows one to calculate cross sections for other measured processes.
As an example, a particular angle-dependent x-ray transition rate, such as the one for K-shell radiative electron capture (K-REC) is used~\cite{Yue2019,Glorius2019,Shao2016,Stohlker1995}.
%This method is used in this paper for the verification. cross sections to be measured are normalized to
The counts of quasi-free proton-deuteron elastic scattering events have also been adopted in some particular cases~\cite{Moskal2009,Zheng2012,Khreptak2019}.
%In the previous experiments at storage ring, to remove this large uncertainty, a relative normalization method was applied.
%Despite this normalization method provides a way to measure cross sections, the K-REC cross section itself is possibly not always known very accurately~\cite{Eichler2007}.

In this paper, we report a method employing the Rutherford scattering distribution for the determination of the reaction luminosity in a storage ring.
%It utilizes the facts that the beam scattering by target is inevitable during the experiment and the intensity of the scattering distribution highly depends on the beam intensity and effective target density.
At low beam energy in a storage ring, the Rutherford scattering is dominant and the differential cross section is known quite well.
%If the scattering differential cross section itself is known quite well, such as the Rutherford scattering cross sections at low energies, the scattering distribution can be well reproduced in the simulation for a certain luminosity.
By normalizing the simulated scattering distribution to the experimental one, the reaction luminosity can be precisely determined. % with a relative uncertainty less than 5\%.
Here, we take an experiment on $^{124}$Xe$(p,\gamma)$ cross section measurement~\cite{Glorius2019} performed at the Experimental Storage Ring (ESR) in GSI as an example to illustrate the power of this method.

\section{Experiment}
%The experiment was performed at the experimental storage ring (ESR) of GSI, Darmstadt.
Taking into account the revolution frequency of the beam of several hundred kHz, the storage ring can increase the beam utilization efficiency by several orders and has been proved to be a facility that is suitable for cross section measurements~\cite{Mei2015,Yue2019}.
In the experiment performed at ESR, a $^{124}$Xe beam was used to measure the proton capture cross section at low energies. % for hot, explosive scenarios such as supernovae and x-ray binaries.
The experiment and the results have been reported in Ref.~\cite{Glorius2019}.
Here we only give a brief introduction.
%\tred{This work was a part of the preparation to this experiment, though the results are applicable to any reaction measurement on a thin target. (Yuanming: I think putting the sentence here may be not very suitable. I modify the last sentence of this paper, see the red text)}

Firstly, the $^{124}$Xe ions were accelerated to about 100 MeV/u by the UNIversal Linear ACcelerator (UNILAC) and SchwerIonenSynchroton (SIS18), and then were extracted to the transfer beam line, completely stripped, injected and stored in the ESR.
The beam with an intensity of about $10^{6}$ to $10^{7}$ per spill was cooled by the electron cooling system and decelerated to the desired energy of a few MeV/u. % radio-frequency system in ESR  by the linear accelerator and the heavy-ion synchrotron
After that, the internal hydrogen gas target with a diameter of about 5 mm and density reaching about $10^{14}$ atoms/cm$^2$~\cite{Kuhnel2009} was switched on.
The beam passed through the gas target with a revolution frequency of about 300 kHz.
The $(p,\gamma)$ reaction products $^{125}$Cs as well as the scattered beam ions were detected with a double-sided silicon strip detector (DSSSD) mounted inside the dipole behind the gas target.

\begin{figure}[htb]
\centering
	\includegraphics[angle=0,width=7.9 cm]{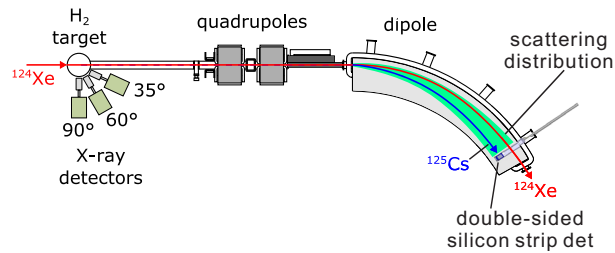}
	\caption{(Colour online) Schematic view of the experimental setup at the ESR from the gas target to the next dipole magnet. %The bending angle of the detector in dipole is 53.5$^{\circ}$.
		\label{ExpSkematic}}
\end{figure}
Figure~\ref{ExpSkematic} shows the schematic illustration of the setup at the ESR including the detector system and the internal gas target.
%The bending angle of the DSSSD detector in the dipole is about 53.5$^{\circ}$. % (see Fig~\ref{ExpSkematic}).
The active area of the DSSSD detector was 4.95 cm $\times$ 4.95 cm, with 16 strips on the front side and 16 perpendicular strips on the back side.
%All the $^{125}Cs$ products as well as part of the scattered $^{124}Xe$ taken as a scattering background can be detected.
Surrounding the gas target, there were installed three high purity germanium semiconductor detectors.
They were used to detect the emitted x-rays from the reaction zone covering angles of 35$^\circ$, 60$^\circ$ and 90$^\circ$ with respect to the beam direction.

\section{Simulation of the scattering distribution}
In the experiment a large part of the ions scattered off the target were measured by the DSSSD detector.
To reproduce the scattering distribution on the detector, the Monte-Carlo (MC) code MOCADI~\cite{Iwasa1997} has been used to simulate the transport of reaction products through the ion-optical system. %Scattering kinematics, ion optical elements (such as quadrupole and dipole magnets) as well as the differential cross sections which determine the scattering angular distribution were all considered.

In the simulation, according to Ref.~\cite{Steck2004}, an estimated beam emittance $\epsilon$=0.5 $mm\times mrad$ and a relative momentum spread $\delta p/p=1\times 10^{-4}$ were adopted. In addition, four different beam energies 5.47 MeV/u, 5.95 MeV/u, 6.65 MeV/u and 6.96 MeV/u have been used. % (there is also a beam energy of 7.92 MeV/u not considered in this paper because its scattering background is not clearly identified).
%The Coulomb barrier for $^{124}Xe$ and $p$ is about 8.96 MeV.
%These energies are just below or close to the threshold energy for non-Rutherford scattering~\cite{Nurmela1998}.
These energies are just around the non-Rutherford threshold energy ($\sim$6.6 MeV/u) predicted by Ref.~\cite{Nurmela1998} for large scattering angle around 165$^\circ$. However, considering: (1) most of the ions detected by the DSSSD detector have much smaller scattering angles well below 100$^\circ$ and (2) the ion scattering through a smaller scattering angle has higher threshold energy, these energies are roughly in a domain where Rutherford scattering is dominant.
%These energies are in a domain where non-Rutherford scattering is negligible~\cite{Nurmela1998}.
Thus, the pure Rutherford scattering was presumed in the simulation.
However, computational efficiency for these simulations is challenging, because of the steepness of the cross section which spans many orders of magnitude for range of scattering angles covered in the experiment.
In this study, MOCADI was only used to simulate scattering kinematics and beam optics adopting a uniform angular distribution.
%However, noting that most of the scattering angles were quite small, to promote computational efficiency, an uniform angular distribution instead of the Rutherford distribution was adopted in the MOCADI simulation.
As a result, a model distribution of the scattered ions on the detector was obtained.
%Of course, the reaction kinematics and the beam optics have already been considered and the information of the initial scattering angle is also maintained.
Afterwards a transformation of the distribution by adding proper weights to the events was introduced to obtain a correct scattering distribution. % using the initial and realistic angular distribution.% (differential cross section).

\begin{figure}[htb]
\centering
	\includegraphics[angle=0,width=7.9 cm]{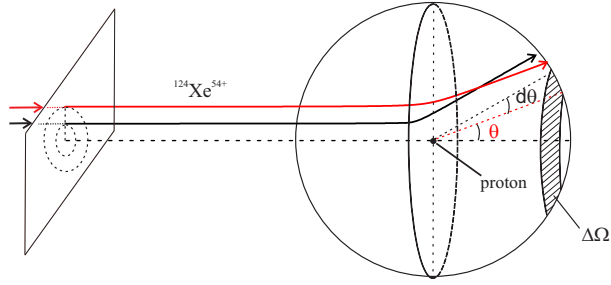}
	\caption{(Colour online) The schematic view of the scattering kinematics. The red and black solid lines represent the track lines of two $^{124}Xe^{54+}$ ions with scattering angel $\theta$ and $\theta+d\theta$, respectively.
		\label{Scattering}}
\end{figure}
A schematic view of the scattering kinematics is shown in Fig.~\ref{Scattering}.
In the experiment, the target is $\rm{H_2}$. However, compared with the beam energy, the binding energy of hydrogen atoms and electrons is negligible. So the proton is actually used as the target in our simulation.
For the solid angle $\Delta\Omega=2\pi sin\theta d\theta$, presented with the shadowed area, the corresponding cross section is $\sigma(\theta)=\frac{d\sigma}{d\Omega}(\theta)\Delta\Omega$, where $\frac{d\sigma}{d\Omega}(\theta)$ is the Rutherford differential cross section:
\begin{equation}\label{Rf}
\frac{d\sigma}{d\Omega}(\theta)=\Big(\frac{1}{4\pi\varepsilon_0}\frac{Z_1Z_2e^2}{4E}\Big)^2\frac{1}{sin^4(\frac{\theta}{2})}.
\end{equation}
Here, $E$ is the kinetic energy in centre of mass system, $Z_1$ and $Z_2$ are the atomic numbers of the target and projectile nuclei respectively.
If the number of the scattered ions within the solid angle $\Delta\Omega$ is $M$ (M>0), a definition of the weight $w(\theta)$ is
 \begin{equation}\label{w}
 w(\theta)=\sigma(\theta) Lt/M.
 \end{equation}
with $L$ being the average luminosity during the experimental time $t$.
It is straightforward, that $w(\theta)$ is the scaling factor to be applied to each ion in the simulation within scattering angles $\theta$ and $\theta+d\theta$.
%In other words, $w$ is the weighting number for the particles in the imulation.
By introducing the corresponding $w$ for all ions in the simulation, the model scattering distribution is transformed into the realistic Rutherford one. % based on the given differential cross section $\frac{d\sigma}{d\Omega}(\theta)$.
%However, the calculation of $w(\theta$ highly depends on the differential cross section.
In our simulation, the $d\theta$ was set to be 0.01 rad.

\section{Luminosity determination}
To determine $L$ and its uncertainty, the maximum likilihood estimation and the $\chi^2$ minimization method have been used~\cite{Hauschild2001}.
Since the two methods gave quite similar results, here we present only the results of the $\chi^2$ minimization method.
\begin{figure}[htbp]
\centering
	\includegraphics[angle=0,width=8.0 cm]{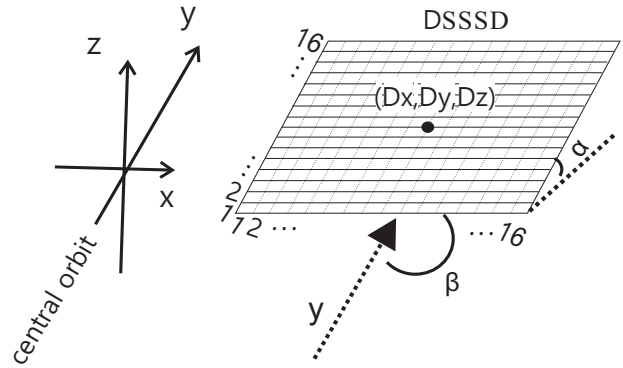}
	\caption{ The geometrical layout of the DSSSD detector. The detector is equipped with 16$\times$16 silicon strips. The central point of the detector is set as (Dx,Dy,Dz). The two angles determining the detector orientation are defined as $\alpha$ and $\beta$. Normally, to get the maximized effective detecting area, both $\alpha$ and $\beta$ would be set to $90^\circ$ in the experiment. However, $\alpha$ was set about $45^\circ$ due to the limited size of the vacuum pipe where the detector inside the dipole magnet was installed. %Dy is restricted by the 53.5$^{\circ}$ bending angle, as shown in Figure ~\ref{ExpSkematic}.
}
 \label{Geometry}
\end{figure}
\begin{figure*}[htbp]
\centering
	\includegraphics[angle=0,width=15 cm]{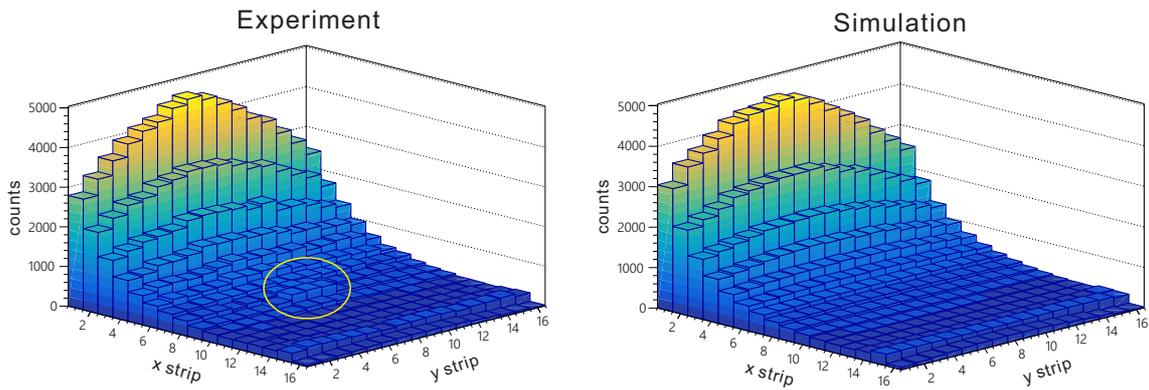}
	\caption{ (Colour online) Left: The experimental scattering distribution detected by the DSSSD detector for the beam energy 5.95 MeV/u. The small region marked with the yellow circle indicates the peak from the $(p,\gamma)$ products. Right: The corresponding simulated scattering distribution on the detector. %with luminosity $2.03\times10^{24}cm^{-2}s^{-1}$.
	\label{Exp_Sim_Default}
}
\end{figure*}

Since the DSSSD detector consists of two sets of 16 silicon strips perpendicular to each other, the detector area is divided into $16\times16$ bins.
We use $i$ and $j$ as the two-dimensional index of the bins ($1\leq i \leq 16$, $1\leq j \leq 16$) and the $\chi^2$ is defined as:
\begin{equation}\label{Dev}
 \chi^2 =\sum_i\sum_j\frac{(N_{ij}^{exp}-N_{ij}^{sim})^2}{\sigma_{ij}^2},
\end{equation}
where $N_{ij}^{sim}$ and $N_{ij}^{exp}$ are the ion numbers detected in the bin ($i, j$) from simulation and experiment, respectively,
$\sigma_{ij}$ is the uncertainty of $(N_{ij}^{exp}-N_{ij}^{sim})$ which was taken as $\sqrt{N_{ij}^{sim}}$~\cite{Hauschild2001}.

If $N$ is the total number of bins considered in the $\chi^2$ calculation and $M$ is the number of free parameters in the simulation, the minimum $\chi^2$, $\chi_{min}^2$, obeys the standard chi-square distribution with $N-M$ degrees of freedom~\cite{Press1989}.
Since all free parameters are adjusted at the same time, the luminosity is determined when $\chi^2$ reaches $\chi_{min}^2$.
The goodness-of-fit test can easily be done by calculating the $Q$ value, which is the probability that the observed $\chi^2$ exceeds the $\chi_{min}^2$ with $N-M$ degrees of freedom~\cite{Hauschild2001}.

In the simulation, besides the luminosity, various parameters have been used, such as the beam parameters including the initial beam energy, emittance and momentum spread, the magnetic fields of the quadrupole and dipole magnets, the detector parameters including the detector position and space orientation, etc.
It is not realistic to keep all these parameters free due to the limitation of computer capabilities.
%Nevertheless, some of the parameters are known as the setup value and some of them can only be roughly estimated.
Nevertheless, a sensitivity test to check the importance of these parameters in a reasonable range is meaningful.
In the test, all parameters were initially set to best estimated values.
Then a chosen parameter for the test was varied and the luminosity value corresponding to the minimum $\chi^2$ value was recorded.
In this way, the sensitivity and importance of the parameters for the luminosity determination was checked.
%%the relation of the determined luminosity with this chosen parameter is roughly built and
For example, the beam emittance was increased by 3 times or the energy of the beam was varied by $\pm20$ keV (the expected energy uncertainty) compared with the set value.
No effect (less than $1\%$) on the luminosity determination has been observed.
Thus, the beam energy was fixed in the simulation to a nominal value given by the electron cooling system.
For many other parameters, similar behavior has been observed.
%the emittance of the beam is varied by 3 times and no effect (less than $1\%$) on the luminosity determination is observed.
%the initial x position of the beam is important in the determination of the minimum $\chi^2$ value
Finally, the geometrical orientation of the DSSSD detector which was not well known in the experiment was found to play the key role in the final luminosity determination.

Figure~\ref{Geometry} shows the specific geometry of the detector as used in the simulation.
%The direction of the beam along the central orbit is defined as y axis.
%The horizontal and vertical direction perpendicular to y axis is defined as x and z axis, respectively.
The coordinates of the detector center were defined as (Dx,Dy,Dz), where the x, y, z directions are the direction pointing to the inner side of the dipole, along the central orbit of the beam and vertically upward, respectively.
%%%Particularly, Dy has been restricted by the 53.5$^{\circ}$ bending angle of the DSSSD detector in the dipole, as shown in Fig.~\ref{ExpSkematic}.
%Dx and Dz can be estimated by the comparison of the $(p,\gamma)$ peak from experiment and MOCADI simulation respectively.
For the orientation of the detector, tilting angles $\alpha$ and $\beta$ were defined as shown in Fig.~\ref{Geometry}.
%Although $\alpha$ and $\beta$ are expected to be around 45$^{\circ}$ and 90$^{\circ}$ respectively in the experiment,
%As shown in Figure.~\ref{Geometry}, there are totally 5 parameters related to the detector geometry setup .
In the simulation, together with the luminosity, these parameters were taken as free parameters.

%%$\chi^2$ minimization is a useful means for estimating parameters even if the measurement errors are not normally distributed

To find $\chi_{min}^2$, the experimental scattering distribution without contributions of other reaction channels is needed.
In the present experiment, the $(p,\gamma)$ products were distributed around the centre of the DSSSD detector (see the bins in the yellow circle in Fig.~\ref{Exp_Sim_Default})~\cite{Glorius2019} and had to be excluded from the calculation.
\iffalse
For the special case of 5.47 MeV/u and 6.65 MeV/u, there are three inner strips close to the beam destroyed during the experiment.
These strips are also excluded in the calculation.
For the particular case of 6.96 MeV/u, the energy is just above the neutron-emission threshold $S_n$=6.71 MeV/u~\cite{Glorius2019}.
In spite of the low $(p,n)$ cross section predicted by theory, the mixing of $(p,n)$ products in the measured Rutherford scattering distribution may still impact the final calculation of $\chi^2$.
\fi

Eventually, a reduced chi-square $\chi_{red}^2 = \chi_{min}^2/(N-M)$ of 1.24, 1.13, 1.11 and 1.38, was obtained for the beam energies of 5.47 MeV/u, 5.95 MeV/u, 6.65 MeV/u and 6.96 MeV/u, respectively.
The corresponding Q values are 0.03, 0.13, 0.19 and 0.01.
According to Ref.~\cite{Press1989}, a model is roughly acceptable if Q>0.001, so the Q values show the credibility of the simulation. As an example, a qualitative comparison of the measured and modelled scattering distributions for the beam energy of 5.95 MeV/u is shown in Fig.~\ref{Exp_Sim_Default}.
\begin{table*}[htp]
 %\hspace{0.5cm}
 \newcommand{\tabincell}[2]
 {\begin{tabular}{@{}#1@{}}#2\end{tabular}}
 \centering
 \fontsize{8.5}{16}\selectfont
 \renewcommand\arraystretch{0.9}
 \begin{threeparttable}
 \caption{The luminosities and the errors determined by the x-ray measurements ((from three independent x-ray detectors at different observation angles as shown in Fig.~\ref{ExpSkematic}) and this work. The last column shows the relative deviation defined as the ratio between the luminosity difference $L-\overline{L}_K$ and the combined error $\delta L$ which is the root-mean-square of the errors from $\overline{L}_K$ and $L$.}
 \label{DataSummary}
 \begin{tabular}{ccccccccc}
 \toprule
 \multirow{2}{*}{\tabincell{c}{\\Energy\\$[MeV/u]$}}&
 \multirow{2}{*}{\tabincell{c}{\\time\\$[s]$}}&
 \multicolumn{1}{c}{ 90$^\circ$}&\multicolumn{1}{c}{ 60$^\circ$}&\multicolumn{1}{c}{ 35$^\circ$}&
 \multirow{2}{*}{\tabincell{c}{Average \\$\overline{L}_K$\\$[\rm{barn/s}]$}}&
 \multirow{2}{*}{\tabincell{c}{This work\\$L$\\$[\rm{barn/s}]$}}&
 \multirow{2}{*}{\tabincell{c}{Relative deviation\\$(L-\overline{L}_K)/\delta L$\\}}\cr
 \cmidrule(lr){3-3} \cmidrule(lr){4-4} \cmidrule(lr){5-5}
 & &\tabincell{c}{$L_K$\\$[\rm{barn/s}]$}&\tabincell{c}{$L_K$\\$[\rm{barn/s}]$}&\tabincell{c}{$L_K$\\$[\rm{barn/s}]$}\cr
 \midrule
 5.47&39120&1.41(0.11)&1.44(0.08)&1.44(0.09)&1.43(0.05)&1.48(0.1)&0.45\cr
 5.95&31544&1.78(0.14)&1.81(0.10)&...&1.80(0.08)&1.98(0.09)&1.4\cr
 6.65&6772&2.90(0.24)&2.89(0.15)&2.83(0.18)&2.87(0.10)&2.93(0.4)&0.15\cr
 6.96&30932&1.79(0.14)&1.83(0.10)&...&1.82(0.08)&1.93(0.3)&0.35\cr
 \bottomrule
 \end{tabular}
 \end{threeparttable}
 \end{table*}

\begin{figure}[htb]
\centering
	\includegraphics[angle=0,width=8 cm]{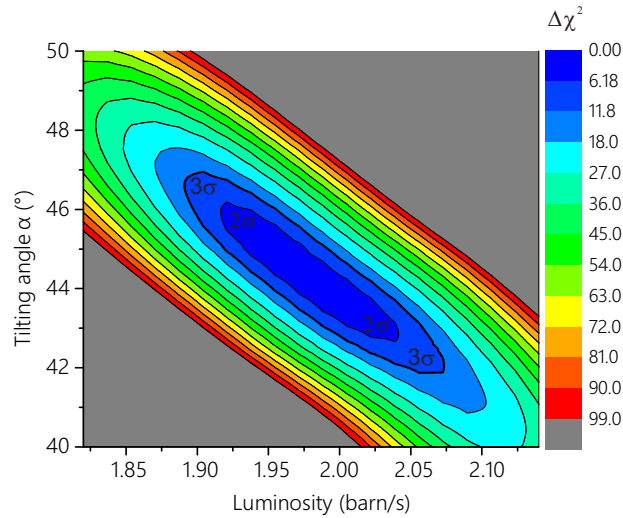}
	\caption{(Colour online) The variation of $\Delta \chi^2$ with tilting angle $\alpha$ and luminosity for the beam energy 5.95 MeV/u.
%The luminosity determination based on the $\chi^2$ statistic method for beam energy 5.95 MeV/u.
The 3$\sigma$ uncertainty ($\Delta \chi^2=11.8$) is adopted for the error estimation and shown with black solid line.
The determined tilting angle $\alpha$ within 3$\sigma$ uncertainty is in agreement with the expected value of $45.0^\circ$.
%The determined tilting angle $\beta$ \iffalse with 3$\sigma$ uncertainty is $90.0^\circ\pm4.4^\circ$ which \fi is in coincidence with the designed value $90^\circ$. %with luminosity $2.03\times10^{24}cm^{-2}s^{-1}$. $1.98(0.10)\times10^{28}cm^{-2}s{-1}$
%Here for the safe side,
		\label{6MeV_L_Dx_x2}}
\end{figure}

Since the parameters (including the luminosity) controlling the simulated Rutherford scattering distribution which is used for the $\chi^2$ calculation are coupled in the simulation, to determine the uncertainty of the extracted luminosity, the following approach has been used.
If $\nu$ parameters from the total of $M$ free parameters ($\nu< M$) are fixed and the remaining $M-\nu$ parameters are varied to minimize $\chi^2$, this minimum value is called $\chi_{\nu}^2$ ($\chi_{\nu}^2>\chi_{min}^2$).
As shown in Ref.~\cite{Press1989}, $\Delta\chi^2\equiv\chi_{\nu}^2-\chi_{min}^2$ is distributed as a chi-square distribution with $\nu$ degrees of freedom.
This connects the projected $\Delta\chi^2$ region with the confidence interval.
For example, for $\nu=2$, $\Delta\chi^2<2.30$ occurs 68.3\% (corresponds to $1\sigma$ for normal distribution), $\Delta\chi^2<6.18$ occurs 95.4\% ($2\sigma$), $\Delta\chi^2<11.8$ occurs 99.73\% ($3\sigma$).

Figure~\ref{6MeV_L_Dx_x2} shows an example of the determination of the luminosity uncertainty.
Here the luminosity and the tilting angle $\alpha$ which is defined in Fig.~\ref{Geometry} are chosen as the fixed parameters ($\nu=2$).
The figure shows the variation of $\Delta \chi^2$ as a function of them.
As stated above, the luminosity uncertainties can be determined when $\Delta \chi^2$ is less than a certain value.
For example, the 2$\sigma$ and 3$\sigma$ uncertainties are determined when $\Delta \chi^2$ is 6.18 and 11.8, respectively.
%To be on the safe side, a 3$\sigma$ uncertainty is adopted as the estimated error,
In this work, we adopted a conservative approach by choosing 3$\sigma$ uncertainty.
It is shown with the black solid line in Fig.~\ref{6MeV_L_Dx_x2}.
%\tred{The contour line of 3$\sigma$ uncertainty is shown with the bold black circles in Fig.~\ref{6MeV_L_Dx_x2}}.
The determined $\alpha=44.3^\circ(2.5^\circ)$ is consistent with the experimental arrangement, in which the angle is expected to be around 45$^\circ$.
From the figure, it is seen that the luminosity determination is very sensitive to $\alpha$.
%The former one is more sensitive.
This is reasonable because the titling angle $\alpha\approx45^\circ$ and if it is slightly changed, the effective area of the DSSSD detector projected on the z direction is considerably changed.
If in future experiments, the tilting angle $\alpha$ is determined precisely (uncertainty of less than $0.5^\circ$), the uncertainty of the luminosity determination will be significantly improved.

The luminosities and the estimated errors for different beam energies are listed in the penultimate column in Table~\ref{DataSummary}.
We can see that for different beam energies, the luminosity does not change significantly.
However, the corresponding errors differ considerably.
This is mainly because the magnitude of the error is highly dependent on the accumulated statistics which was different in all cases. %intensity or the extent of the experimental scattering distribution and the situation is different for different case.
For example, for the case of 6.65 MeV/u, the experiment time was much shorter than in other cases and the detector performance degraded due to an increased ion dose level.
%In addition, 3 damaged strips of the DSSSD detector complicated the situation.
As a result, the accumulated statistics of scattering events was the lowest and hence the obtained error is the largest.
For another case of 6.96 MeV/u, the yield of $(p,\gamma)$ product is much higher than in other cases mainly because of the largest $(p,\gamma)$ cross section~\cite{Glorius2019}.
Accordingly, the area dominated by $(p,\gamma)$ becomes larger, reducing the effective data set for the Rutherford scattering model.
At this or even higher energy, the $(p,n)$ products and maybe also the non-Rutherford scattering can contribute to the scattering events which increase the luminosity uncertainty.
%All these factors make the determined luminosity error for this energy become larger.

\section{The verification of this method}
As stated in the introduction, the luminosity can also be determined through the K-REC x-ray measurement.
This provides a good opportunity to check the validity of this work.
In this case, the luminosity (noted as $L_K$) can be expressed as
\begin{equation}\label{Lt}
   L_K=\frac{N_{K}}{\epsilon (d\sigma_{K}/d\Omega)\Delta\Omega t}
\end{equation}
where $N_{K}$ is the number of K-REC x-rays, %which are measured by Ge detectors,
$\epsilon$ is the intrinsic efficiency of the Ge detector, $\Delta\Omega$ is the solid angle spanned by the Ge detector and $d\sigma_{K}/d\Omega$ is the theoretical differential K-REC cross section.
The individual efficiency-corrected K-REC counts per steradian $N_{K}/\epsilon\Delta\Omega$ and the effective theoretical $d\sigma_{K}/d\Omega$ can be found in Ref.~\cite{Glorius2019} for each beam energy.
Based on these values and the measurement duration time $t$ listed in Table~\ref{DataSummary}, $L_K$ from three different angles and the average $\overline{L}_K$ were calculated.

The relative deviation between our present results and the K-REC method are listed in the last column of Table~\ref{DataSummary}.
All the deviation relative to the combined error $\delta L$ are positive and well below 2.
This systematic deviation may be induced by some unconsidered factors in the simulation of the experiment.
This method was not foreseen to be used in this experiment~\cite{Glorius2019}.
Thus, since no dedicated efforts were undertaken to determine relevant parameters, the good agreement between two methods is remarkable and indicates the power of this new approach.

%This systematical deviation may be induced by some unconsidered factors from the simulation or experiment, or the potential overestimation of the theoretical K-REC differential cross section~\cite{Eichler2007}.

The averaged errors obtained by the K-REC method are listed in Table~\ref{DataSummary}.
For the cases with high statistics, the errors from the two methods are comparable.
%The error magnitude to the intensity of the scattering background which is a function of multiplication of luminosity and experimental time. However, it's not easy to give a analytic expression of the error.

\section{Summary and outlook}

In the direct measurement of absolute reaction cross sections, the luminosity is a critical quantity, which is hard to determine precisely.
This is especially true for experiments using thin gas targets.
%Also, the scattering distribution in the experiment usually has to be blocked or subtracted in the measurement.
Just like the x-ray emissions, the scattering distribution itself reflects the collision frequency and is useful for the determination of the luminosity.

As a proof of concept, this work has shown the feasibility to use the elastic scattering distribution to precisely obtain the luminosity.
A weighting method is introduced for the simulation of the scattering distribution on the detector. % we adopted a MC code MOCADI for the Rutherford scattering simulation and a weighting method to improve the calculation efficiency.
By taking the $^{124}$Xe + H$_{2}$ experiment performed at the ESR storage ring as an example~\cite{Glorius2019}, the luminosity is determined via a $\chi^2$ minimization approach.
Although, a small systematic offset to the established K-REC method is found, there is still agreement within the error bars demonstrating the validity of the method.
%Although there seems to be a systematic deviation when compared with the K-REC x-ray measurements, still an agreement within the error bars indicates the validity of the present method. %using the scattering background as a tool to characterize the luminosity.
The uncertainties determined by the two independent and complementary methods are comparable.

This new method will become indispensable in the future experiments as planned in the dedicated low-energy storage rings~\cite{Lestinsky2016,Grieser2012} with light beams at much lower energies, e.g., to reach the Gamow window for astrophysical reaction studies~\cite{Rauscher2010}.

In the future, with a precisely defined geometry of the detector, the uncertainty of the determined luminosity is expected to be significantly reduced. %, which will be comparable with the luminosity uncertainty determined from the K-REC X-ray measurement.
Furthermore, if the measurement of the scattering distribution can be performed just before the dipole, the simulation would be simplified as the transport of the ions through magnetic system is not needed, and the determination of the luminosity will become more accurate.

It should be mentioned that when the beam energy is high enough, the non-Rutherford scattering should be accounted for and the optical model can be used for the calculation of the corresponding differential cross section~\cite{Koning2003}.
However, with the higher energy, the reaction products from nuclear channels like $(p,n)$ and $(p,\alpha)$ may mask scattering events.
In such situation, a target recoil measurement close to the gas target can be considered~\cite{Yue2019,Zamora2017}.

%Although this method is based on the experiment performed at a storage ring, it can also be employed at facilities where the luminosity needs to be precisely measured.
Although this method is based on the experiment performed at a storage ring, it can also be employed to any reaction measurement on a thin target, where the luminosity needs to be precisely measured.

\section*{Acknowledgments}

This work is supported in part by the European Research Council (ERC) under the European Union's Horizon 2020 research and innovation programme (Grant Agreement No. 682841 "ASTRUm"), the NSFC (Grants No. 11905261), the Key Research Program of Frontier Sciences of CAS (Grant No. QYZDJ-SSW-S), the National Key R\&D Program of China (Grant No. 2016YFA0400504 and No. 2018YFA0404400) and the Helmholtz-CAS Joint Research Group (Grant No. HCJRG-108). Y.M.X. thanks for support from CAS "Light of West China" Program. Y.A.L. acknowledges support by the CAS President's International Fellowship Initiative (Grant No. 2016VMA043).
%Y.H.Z. acknowledges support by the ExtreMe Matter Institute EMMI at the GSI Helmholtzzentrum f{\"u}r Schwerionenforschung, Darmstadt, Germany.

\end{document}